\documentclass[aps,showkeys,superscriptaddress,preprint,amsmath,amssymb,showpacs]{revtex4}%,showpacs
\usepackage{graphicx,rotating}% Include figure files
\usepackage{dcolumn}% Align table columns on decimal point
\usepackage{color}
\usepackage{float}

\begin{document}

\title{Tricritical point and phase diagram based on critical scaling in monoaxial chiral helimagnet Cr$_{1/3}$NbS$_{2}$}
%\title{Magnetic properties in chiral helimagnet Cr$_{1/3}$NbS$_{2}$ single crystal}

\author{Hui Han}
\affiliation{Anhui Key Laboratory of Condensed Matter Physics at Extreme Conditions, High Magnetic Field Laboratory,
Chinese Academy of Sciences, Hefei 230031, China}\affiliation{University of Science
and Technology of China, Hefei 230026, China}
\author{Lei Zhang}
\email[Corresponding author. Email:]{zhanglei@hmfl.ac.cn}
\affiliation{Anhui Key Laboratory of Condensed Matter Physics at Extreme Conditions, High Magnetic Field Laboratory,
Chinese Academy of Sciences, Hefei 230031, China}%; Tel.:+86 551
%65595141; fax:+86 551 65591149
\author{Deepak  Sapkota}
\affiliation{Department of Physics and Astronomy, The University of Tennessee, Knoxville, Tennessee 37996, USA}
\author{Ningning Hao}
\author{Langsheng Ling}
\author{Haifeng Du}
\affiliation{Anhui Key Laboratory of Condensed Matter Physics at Extreme Conditions, High Magnetic Field Laboratory,
Chinese Academy of Sciences, Hefei 230031, China}
\author{Li Pi}
\affiliation{Anhui Key Laboratory of Condensed Matter Physics at Extreme Conditions, High Magnetic Field Laboratory,
Chinese Academy of Sciences, Hefei 230031, China} \affiliation{Hefei National
Laboratory for Physical Sciences at the Microscale, University of
Science and Technology of China, Hefei 230026, China}
\author{Changjin Zhang}
\affiliation{Anhui Key Laboratory of Condensed Matter Physics at Extreme Conditions, High Magnetic Field Laboratory,
Chinese Academy of Sciences, Hefei 230031, China}

\author{David G. Mandrus}\email[Corresponding author. Email:]{dmandrus@utk.edu}
\affiliation{Department of Physics and Astronomy, The University of Tennessee, Knoxville, Tennessee 37996, USA}
\affiliation{Materials Science and Technology Division, Oak Ridge National Laboratory, Oak Ridge, Tennessee 37831, USA}
\affiliation{Department of Materials Science and Engineering, The University of Tennessee, Knoxville, Tennessee 37996, USA}

\author{Yuheng Zhang}
\affiliation{Anhui Key Laboratory of Condensed Matter Physics at Extreme Conditions, High Magnetic Field Laboratory,
Chinese Academy of Sciences, Hefei 230031, China} \affiliation{Hefei National
Laboratory for Physical Sciences at the Microscale, University of
Science and Technology of China, Hefei 230026, China}

\date{\today}

\begin{abstract}

In this work, the magnetism of the single crystal Cr$_{1/3}$NbS$_{2}$, which exhibits chiral magnetic soliton lattice (CSL) state, is investigated.
The magnetization displays strong magnetic anisotropy when the field is applied perpendicularly and parallel to the $c$-axis in low field region ($H<H_{S}$, $H_{S}$ is the saturation field).
The critical exponents of Cr$_{1/3}$NbS$_{2}$ are obtained as $\beta=$ 0.370(4), $\gamma=$ 1.380(2), and $\delta=$ 4.853(6), which are close to the theoretical prediction of three-dimensional Heisenberg model.
Based on the scaling equation and the critical exponents, the $H-T$ phase diagram in the vicinity of the phase transition is constructed, where two critical points are determined. One is a tricritical point which locates at the intersection between the CSL, forced ferromagnetic (FFM), and paramagnetic (PM) states. The other one is a critical point situated at the boundaries between CSL, helimagnetic (HM), and PM states.

\end{abstract}


%75.30.Gw Magnetic anisotropy

%75.40.-s: Critical-point effects, specific heats, short-range
%order\\

%75.40.Cx: Static properties (order parameter, static
%susceptibility, heat capacities, critical exponents, etc.)\\

%75.40.Gb-Dynamic properties (dynamic susceptibility, spin waves,
%spin diffusion, dynamic scaling, etc.)


%Ferromagnetic, 4673 %Acta Mat
%
%Single crystal, 3439, 3887
%
%Anisotropy, 1671, 2476, 862

\maketitle

\section{Introduction}

Magnetic materials with chirality have attracted considerable attention due to spin-textures such as helimagnetic (HM) structure, conical magnetic ordering, magnetic skyrmion, chiral bobber \cite{Robler2,Muuhlbauer,YuNature,Seki,Rybakov}.
The chirality with spin-orbital coupling in the crystal structure results in an antisymmetric exchange interaction called the Dzyaloshinskii-Moriya (DM) interaction, which is 1$\sim$2 orders of magnitude weaker than the ferromagnetic coupling \cite{Dzyaloshinsky,Moriya}.
The competition between the DM interaction with the ferromagnetic coupling causes the appearance of chiral spin-texture \cite{Bauer}.
When an external magnetic field is applied above a threshold value, the HM ordering is usually modulated into particle-like spin-texture such as skyrmion, magnetic soliton.

Recently, the monoaxial chiral magnet Cr$_{1/3}$NbS$_{2}$ becomes prominent due to the chiral magnetic soliton lattice (CSL), which is a type of superlattice structure consist of periodic helical spin texture \cite{Togawa2013,Togawa2012,Masaki}.
The crystal structure for Cr$_{1/3}$NbS$_{2}$ belongs to the space group $P6_{3}22$ \cite{Ghimire2013,Miyadai}.
The hexagonal layers in 2H-NbS$_{2}$ are intercalated by the Cr ions which are in the trivalent state with localized moments $S=3/2$, whereas the electronic conduction originates from an unfilled band of Nb atoms.
Due to the strong magnetocrystalline anisotropy and DM interaction, Cr$_{1/3}$NbS$_{2}$ displays a ground state of helical magnetic ordering with vector along the $c$-axis \cite{Miyadai,Hulliger}.
The magnetic sate can be modulated differently through the direction of the applied external field.
When $H$ is applied parallel to the $c$-axis, the HM ordering is polarized to conical ordering, and finally to forced ferromagnetical (FFM) phase.
However, when $H$ is applied perpendicularly to the $c$-axis, the HM structure with 48 nm changes continuously into CSL state, which is well reproduced by the one-dimensional chiral sine-Gordon model \cite{Togawa2012}.
Further increase of $H$ results in a phase transition from an incommensurate CSL to a commensurate FFM state \cite{Tsuruta}.
It has been demonstrated that CSL can be effectively manipulated by magnetic field or current injection, which supplies potential application as spintronic device \cite{Kishine,Borisov,Koumpouras}.

Although the phase transition and CSL in Cr$_{1/3}$NbS$_{2}$ have been intensively investigated, controversies emerge for the complex phase diagrams \cite{Mankovsky,Sirica,Togawa2016JPSJ,Togawa2015PRB,Chapman}.
The phase diagrams have been constructed in several experimental and theoretical works \cite{Ghimire2013,Tsuruta,Laliena,Nishikawa,Shinozaki,Laliena2016}.
However, vagueness and difference exist on boundaries between different phases, especially in the vicinity of the phase transition.
In this work, the $H-T$ phase diagram of the single crystal Cr$_{1/3}$NbS$_{2}$ is constructed based on the critical scaling method, where the phases and boundaries in the vicinity of the transition are clearly clarified.
The critical behavior unambiguously indicates that the magnetic coupling is of a three-dimensional (3D) Heisenberg type, which suggests that the magnetic coupling of Cr-Cr occurs not only within the $ab$-plane but also between the interlayers along the $c$-axis.
In addition, two critical points are determined on the $H-T$ phase diagram.

\section{experimental methods}

Single crystals of Cr$_{1/3}$NbS$_{2}$ were grown by the chemical vapor transport(CVT) method using iodine as the transport agent \cite{Miyadai}.
The chemical compositions were carefully checked by Energy Dispersive X-ray(EDX) spectrometry.
The crystal structure was confirmed by the Rigaku-TTR3 X-ray diffractometer using high-intensity graphite monochromatized Cu K$\alpha$ radiation.
The magnetization was measured using a Quantum Design Vibrating Sample Magnetometer (SQUID-VSM).
The no-overshoot mode was applied to ensure a precise magnetic field.
The magnetic field was relaxed for two minutes before data collection.
For the measurement of initial isothermal magnetization, the sample was firstly heated adequately above $T_{C}$ for ten minutes, then cooled to the target temperature under zero magnetic field.

\section{Results and discussion}

\begin{figure}%[B]
\includegraphics[width=1.0\textwidth,angle=0]{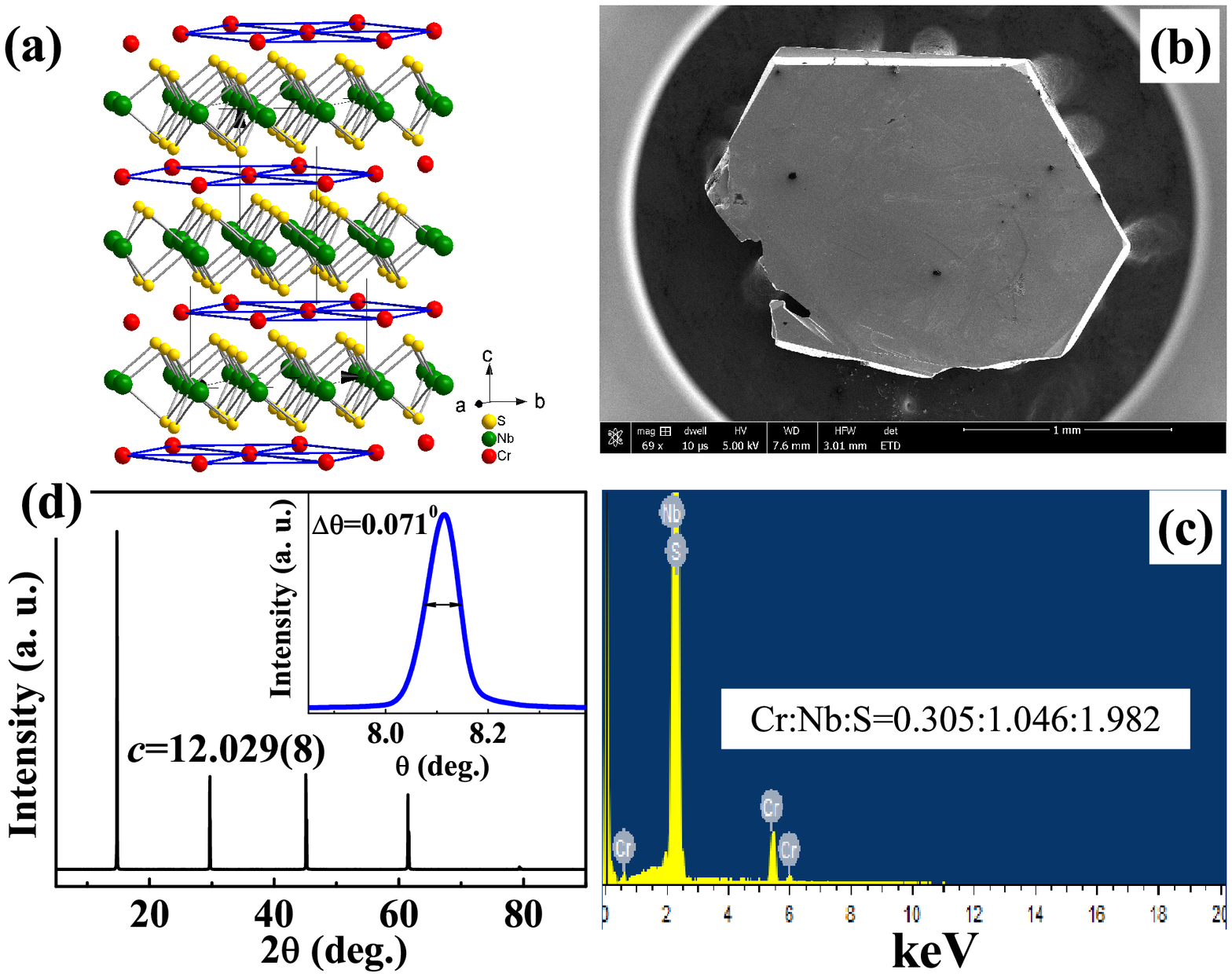}
\caption{(Color online) (a) The crystal structure of Cr$_{1/3}$NbS$_{2}$; (b) the morphology of Cr$_{1/3}$NbS$_{2}$ single crystal;
(c) a typical EDX spectrum for single crystal Cr$_{1/3}$NbS$_{2}$; (d) XRD pattern of the plane (the inset shows the rock curve).} \label{structure}
\end{figure}

Figure \ref{structure} (a) shows the crystal structure for Cr$_{1/3}$NbS$_{2}$.
The NbS$_{2}$ is a quasi-two-dimensional system with van der Waals interaction between the layers.
The Cr atoms are intercalated in the octahedral holes between the trigonal prismatic layers of 2H-NbS$_{2}$ \cite{Ghimire2013}.
Figure \ref{structure} (b) gives the morphology of the Cr$_{1/3}$NbS$_{2}$ single crystal.
The single crystal presents layered characteristics, with a bright hexagonal surface of 1 mm $\times$ 1mm.
The chemical proportion is determined by the EDX spectrum as depicted in Fig. \ref{structure} (c), which gives that the proportion of Cr : Nb : S is close to 0.33 : 1: 2.
The bright surface was checked by XRD as shown in Fig. \ref{structure} (d), which indicates that the surface is (001) plane.
The inset of Fig. \ref{structure} (d) presents the rock curve of the (001) diffraction peak, the full-width-at-half-maximum (FWHM) of which is $\Delta$$\theta$ = 0.071$^{\circ}$.
The single peak and narrow FWHM of the rock curve indicate high quality of the single crystal sample without twin crystal.
The XRD pattern of the single crystal gives the lattice constant $c= 12.029(8)$ $\AA$, which is very close to previous reports \cite{Miyadai,Hulliger,Dyadkin}.

\begin{figure}%[B]
\includegraphics[width=1.0\textwidth,angle=0]{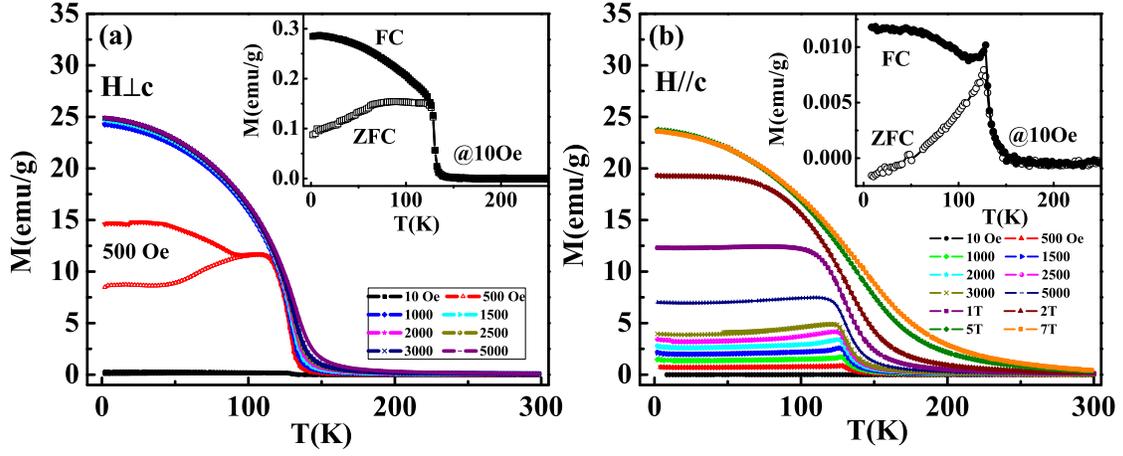}
\caption{(Color online) Temperature dependence of magnetization [$M(T)$] under selected fields for Cr$_{1/3}$NbS$_{2}$ with (a) $H\perp c$ and (b) $H//c$ (the insets show the ZFC and FC curves under 10 Oe with $H\perp c$ and $H//c$ respectively).} \label{mt}
\end{figure}

Figure \ref{mt} (a) and (b) gives the temperature dependence of magnetization [$M(T)$] under selected field, where the external field is applied perpendicularly and parallel to the $c$-axis respectively.
All $M(T)$ curves with $H\perp c$ or $H//c$ undergo a magnetic ordering transition at the phase transition temperature $T_{C}$ $\sim$ 125 K, which is in agreement with previous report \cite{Miyadai}.
However, the $M(T)$ curves exhibit different behaviors when $H\perp c$ and $H//c$ .
When $H\perp c$, $M(T)$ curves below $H=$ 500 Oe almost display the same behavior, as shown in Fig. \ref{mt} (a).
However, when $H$ exceeds 500 Oe, the values under fixed field increase abruptly.
On the other hand, when $H//c$, $M(T)$ curves are raised monotonously with the applied field, as presented in Fig. \ref{mt} (b).
These different behaviors of $M(T)$ between $H\perp c$ and $H//c$ can be attributed to the formation of CSL \cite{Miyadai,Hulliger}.
When $H\perp c$, the CSL forms under the lower field, thus the $M(T)$ curves exhibit similar behaviors in the CSL phase.
However, when under higher field, the CSL is destroyed abruptly, which causes the prominent different behaviors in $M(T)$ curves.
When $H//c$, no CSL phase forms but a conical ordering is polarized gradually.
Therefore, no abrupt change appears in the $M(T)$ curves when $H//c$.
In addition, it is also noticed from the insets of Fig. \ref{mt} that bifurcations occur to the zero-field-cooling (ZFC) and field-cooling (FC) curves, which are attributed to the different magnetic ground state under zero field and non-zero-field.
As we know, the HM is polarized to conical spin ordering by the external field.
In a sequence of ZFC, the system exhibits a HM ordering, where the HM ordering is gradually polarized by the field with the increase of temperature.
However, In a sequence of FC, the system displays a ground state of conical ordering.
Thus, the different magnetic ground states under zero-field and non-zero-field result in the different behaviors between ZFC an FC curves \cite{Miyadai,Hulliger}.

\begin{figure}%[B]
\includegraphics[width=0.8\textwidth,angle=0]{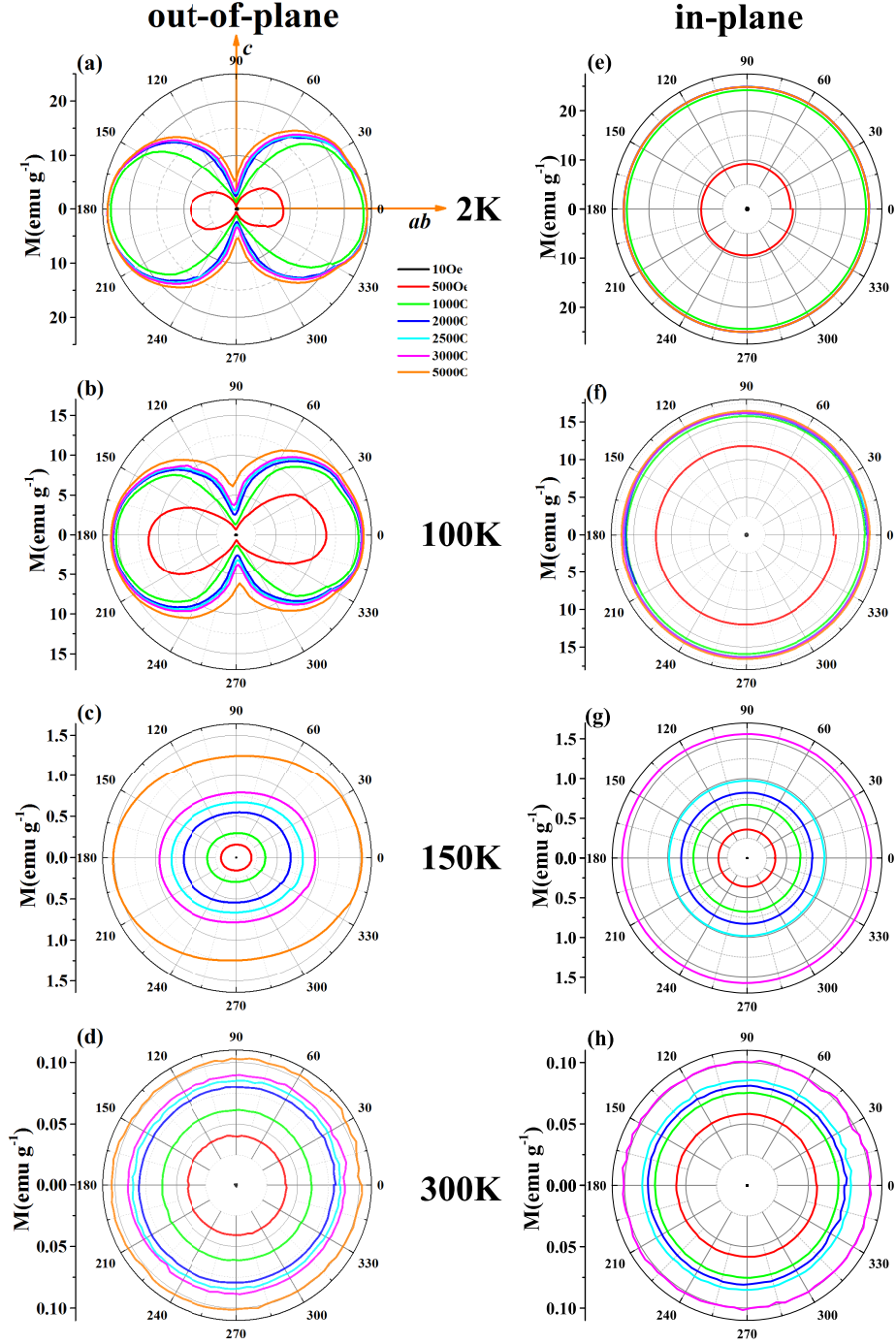}
\caption{(Color online) (a) The out-of-plane and in-plane magnetization as a function of rotation angel [$M(\varphi)$] at selected temperature for Cr$_{1/3}$NbS$_{2}$ (the curves under different fields are distinguished by different colors).} \label{rot}
\end{figure}

As indicated by the $M(T)$ curves, the monoaxial Cr$_{1/3}$NbS$_{2}$ exhibits strong anisotropic magnetization, which should be investigated by the angle dependent of magnetization [$M(\varphi)$].
Figure \ref{rot} shows the out-of-plane and in-plane $M(\varphi)$ at selected temperature.
The in-plane $M(\varphi)$ was measured under the field within the $ab$-plane, while the out-of-plane one was performed under the field rotated from $c$-axis to $ab$-plane.
It can be seen that the in-plane $M(\varphi)$ display isotropic magnetization.
However, out-of-plane $M(\varphi)$ at 2 K and 100 K exhibits a shape of butterfly, which suggests strong magnetic anisotropy.
The magnetization in $ab$-plane is much stronger than that along $c$-axis, suggesting that the spins are ordered within the $ab$-plane.
The out-of-plane $M(\varphi)$ at 150 K exhibits slightly magnetic anisotropy, which is much weaker than those at 100 K or 2 K.
The slight magnetic anisotropy at 150 K may be attributed to the precursor or fluctuation phenomenon analogy to that in FeGe \cite{Wilhelm2011}.
The out-of-plane $M(\varphi)$ at 300 K does not display any magnetic anisotropy.

\begin{figure}%[B]
\includegraphics[width=1.0\textwidth,angle=0]{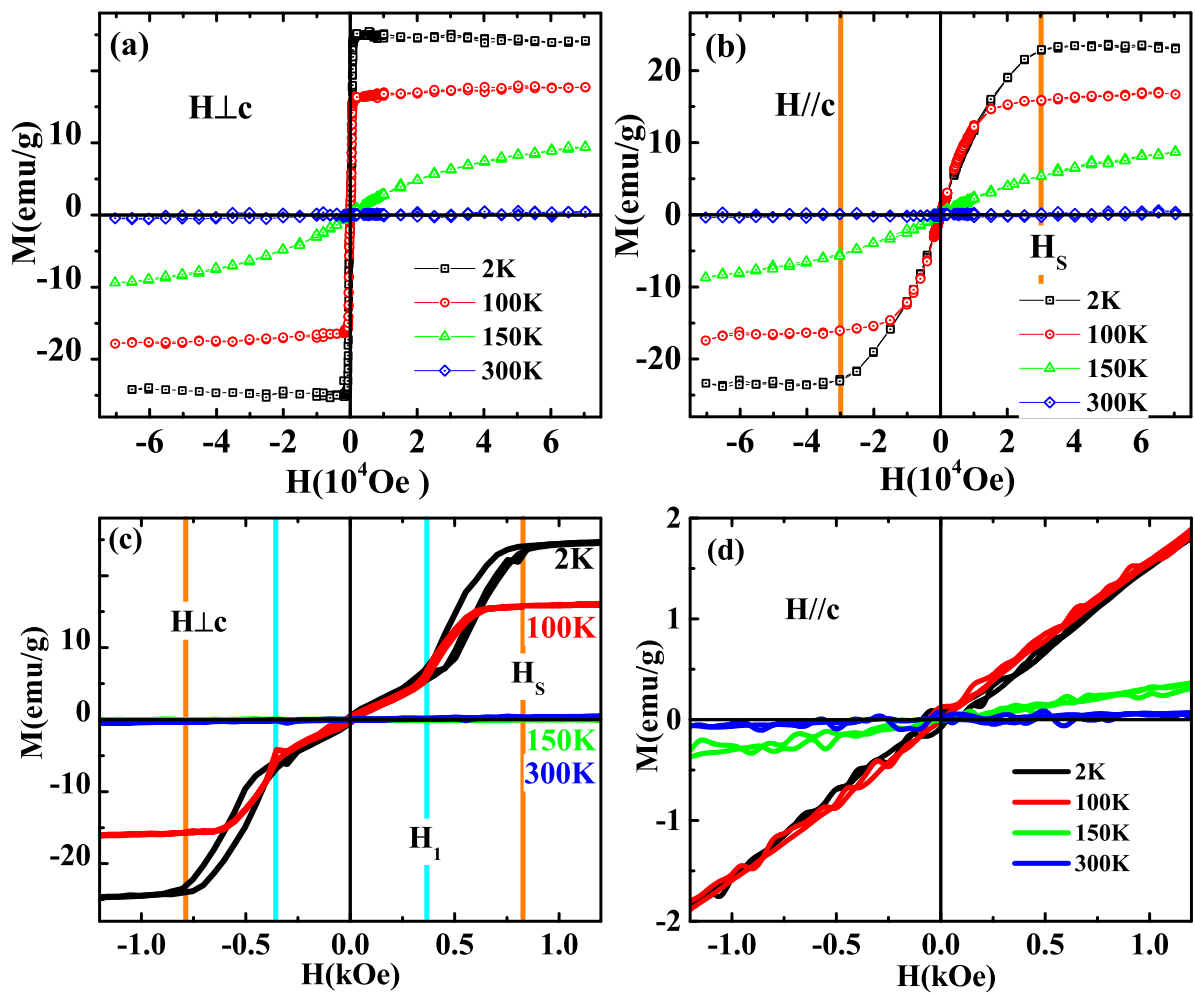}
\caption{(Color online) (a) and (b) Isothermal magnetization as a function of field [$M(H)$] at selected temperature with $H\perp c$ and $H//c$; (c) and (d) the magnified $M(H)$ in low field region.} \label{mh}
\end{figure}

\begin{figure}%[B]
\includegraphics[width=1.0\textwidth,angle=0]{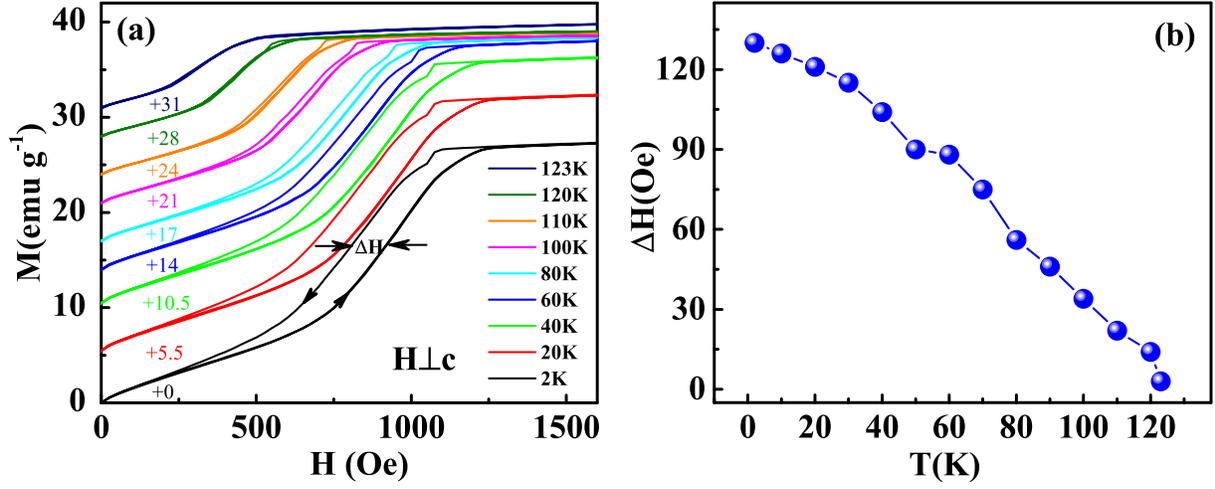}
\caption{(Color online) (a) The $M(H)$ loops with $H\perp c$ (the curves are elevated vertically for clarity, and the elevation values are marked); (b) the width of the $M(H)$ loop ($\Delta H$) as a function of temperature.} \label{mh-loop}
\end{figure}

Figure \ref{mh} (a) and (b) plot the isothermal magnetization [$M(H)$] at selected temperature with $H\perp c$ and $H//c$, respectively.
The magnified $M(H)$ curves in low field region are depicted in Fig. \ref{mh} (c) and (d).
When $H\perp c$, $M$ becomes saturated rapidly with very small saturation field ($H_{S} \sim$ 0.7 kOe), as shown in Fig. \ref{mh} (c).
When $H//c$, $M$ becomes saturated slowly with large saturation field ($H_{S} \sim$ 25 kOe), as given in Fig. \ref{mh} (b).
However, the same saturation magnetization ($M_{S}$) is attained regardless of the direction of applied field.
For $M(H)$ with $H\perp c$ at 2 K and 100 K below $T_{C}$, it exhibits two symmetrical magnetic steps in low field region before saturation.
Moreover, a loop is found at the magnetic step between increase and decrease of field, which has been also found in previous works \cite{Tsuruta2016JPSJ}.
However, when $H//c$, no step or loop is found on the $M(H)$ curves.
To better understand the $M(H)$ loops, the $M(H)$ with $H\perp c$ was measured at different temperatures, as shown in Fig \ref{mh-loop} (a). The curves in Fig \ref{mh-loop} (a) are elevated vertically for clear clarity.
As can be seen, the loop becomes smaller as the temperature increases.
The width of the $M(H)$ loop ($\Delta H$) as a function of temperature is extracted to plot in Fig. \ref{mh-loop} (b), which displays that the $\Delta H$ decreases with the increase of temperature.
The loop disappears when temperature exceeds 120 K.
It is well known that Cr$_{1/3}$NbS$_{2}$ undergoes a field-induced transition from HM to CSL below $T_{C}$.
Moreover, the CSL state is polarized to FFM, accompanied with a transition from incommensurate to commensurate states.
In fact, this kind of $M(H)$ loop is usually found in the magnetic metastable transition system, such as UP$_{0.9}$S$_{0.1}$ and EuSe \cite{Stryjewski}.
The $M(H)$ loop originates during the formation and collapse of the magnetic metastate by the applied field \cite{Stryjewski}.
The $M(H)$ loops in Cr$_{1/3}$NbS$_{2}$ assemble those observed in the ferrimagnetic or antiferromagnetic metastable, where $M(H)$ loop occurs between phase transitions induced by the field.
Therefore, $M(H)$ loops in Cr$_{1/3}$NbS$_{2}$ confirms that the CSL state is a magnetic metastable \cite{Stryjewski}.

\begin{figure}%[B]
\includegraphics[width=1.0\textwidth,angle=0]{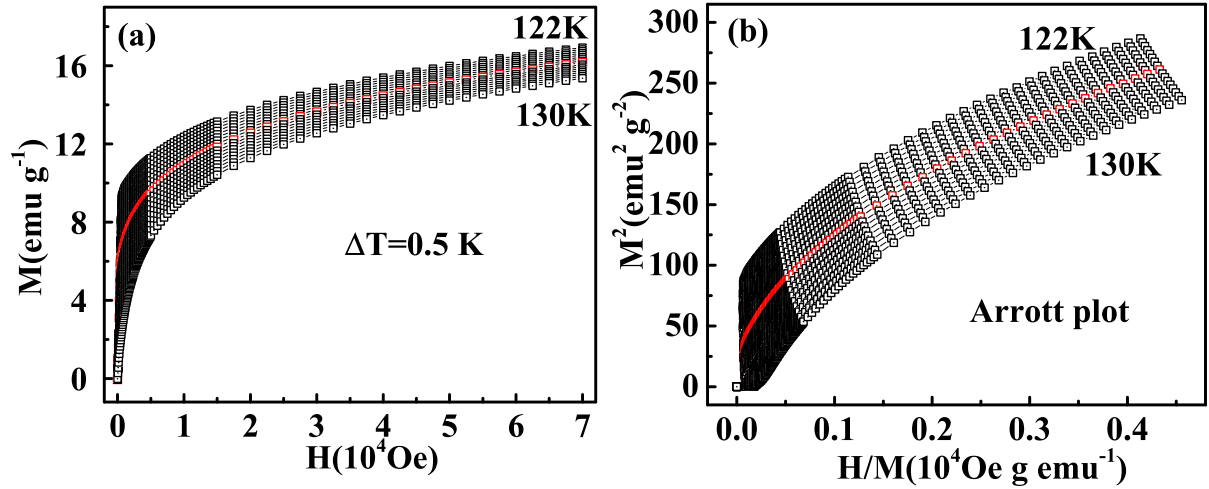}
\caption{(Color online) (a) The initial isothermal $M(H)$ around $T_{C}$ with $H\perp c$; (b) the Arrott plot for Cr$_{1/3}$NbS$_{2}$ (the curves at $T_{C}$ are marked in red color).} \label{Arrott}
\end{figure}

In order to uncover the magnetic coupling in Cr$_{1/3}$NbS$_{2}$, the critical behavior is investigated.
Figure \ref{Arrott} (a) gives the initial isothermal $M(H)$ around $T_{C}$ with $H\perp c$, and the Arrott plot of $M^{2}$ vs. $H/M$ is depicted in Fig. \ref{Arrott} (b).
All curves on the Arrott plot display nonlinear behaviors even in the high field region, which suggests that the magnetic interaction in Cr$_{1/3}$NbS$_{2}$ cannot be described by the conventional Landau mean-field model \cite{Arrott,Kaul}.
The order of the phase transition can be determined by the slope from the Arrott plot according to the Banerjee's criterion, where a negative slope suggests a first-order transition and a positive slope implies a second-order one \cite{Banerjee}.
The positive slopes of $M^{2}$ vs. $H/M$ curves reveal that the phase transition in Cr$_{1/3}$NbS$_{2}$ is of a second order.

\begin{figure}%[B]
\includegraphics[width=1.0\textwidth,angle=0]{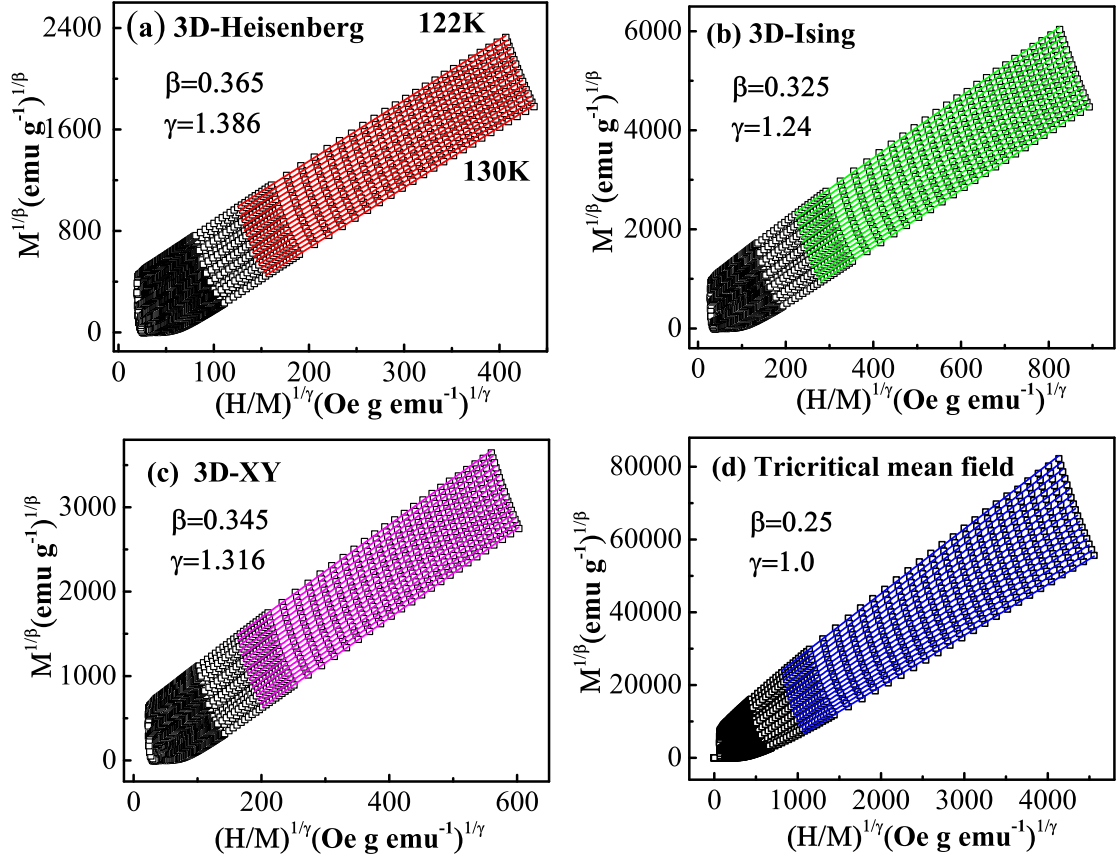}
\caption{(Color online) The isotherms of $M^{1/\beta}$ vs.
$(H/M)^{1/\gamma}$ fitted by lines with parameters of (a) 3D-Heisenberg model (red), (b) 3D-Ising
model (green), (c) 3D-XY model (magenta), and (d) tricritical mean-field model (blue).} \label{MAP}
\end{figure}

\begin{figure*}%[B]
\includegraphics[width=0.8\textwidth,angle=0]{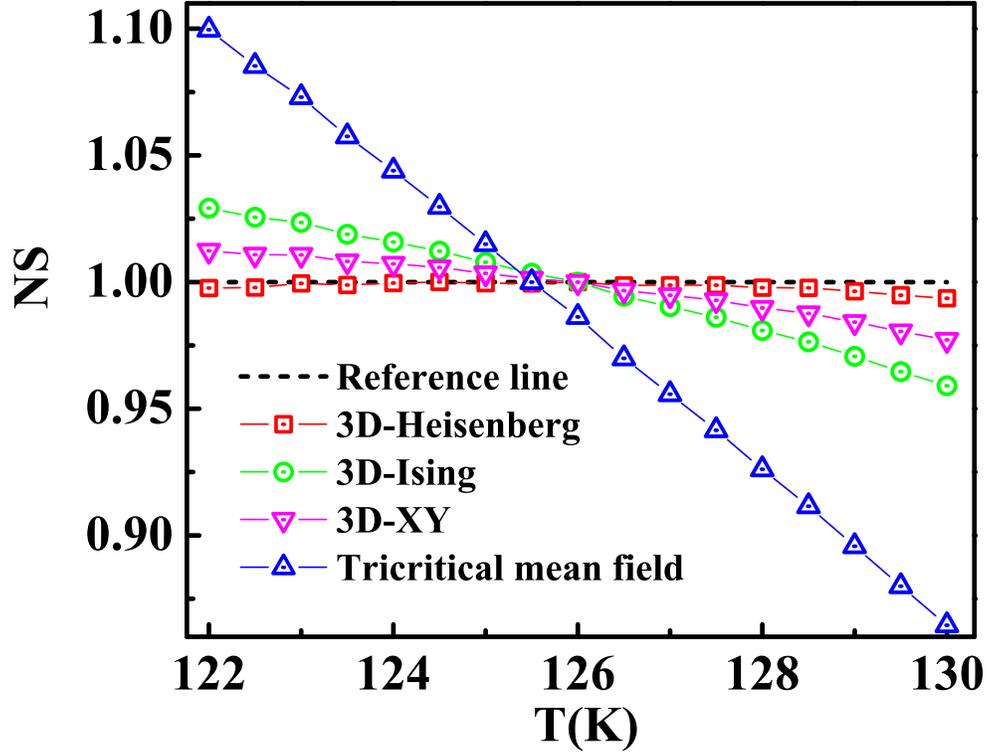}
\caption{(Color online) Temperature dependence of normalized slope ($NS$) for the four theoretical models.} \label{ns}
\end{figure*}

In the vicinity of a second order magnetic phase transition, the spontaneous magnetization $M_{S}$ and initial
susceptibility $\chi_{0}$ can be described by a series of functions \cite{FisherRMP,Stanley}:
\begin{equation}
M_{S}(T)=M_{0}(-\varepsilon)^{\beta}, \varepsilon<0,
T<T_{C}\label{1}
\end{equation}
\begin{equation}
\chi_{0}^{-1}(T)=(h_{0}/M_{0})\varepsilon^{\gamma}, \varepsilon>0,
T>T_{C}\label{2}
\end{equation}
\begin{equation}
M=DH^{1/\delta}, \varepsilon=0, T=T_{C}\label{3}
\end{equation}
where $\varepsilon=(T-T_{C})/T_{C}$ is the reduced temperature;
$M_{0}/h_{0}$ and $D$ are critical amplitudes. The parameters
$\beta$ (associated with $M_{S}$), $\gamma$ (associated with
$\chi_{0}$), and $\delta$ (associated with $T_{C}$) are the
critical exponents. The critical behavior around the critical temperature can be described by a series of critical exponents, which follow the Arrott-Noakes equation of state in asymptotic critical region \cite{Arrott2}:
\begin{equation}
(H/M)^{1/\gamma}=(T-T_{C})/T_{C}+(M/M_{1})^{1/\beta}
\end{equation}
where $M_{1}$ is a constant. The critical exponents implies significant clues of magnetic interactions, such as the correlating length, spin-dimensionality, and decaying distance of magnetic coupling.
Four series of critical exponents belonging to the 3D-Heisenberg model ($\beta= 0.365$, $\gamma=1.386$), 3D-Ising model ($\beta= 0.325$, $\gamma=1.24$), 3D-XY model ($\beta= 0.345$, $\gamma=1.316$), and tricritical mean-field model ($\beta= 0.25$, $\gamma=1.0$) are tried to construct the modified Arrott plots \cite{Kaul,Huang}, as shown in Figs. \ref{MAP} (a), (b), (c), and (d) respectively.
All the curves in these four constructions exhibit quasi-straight lines in high field region.
However, the lines in Fig. \ref{MAP} (d) are not parallel to each other, indicating that the tricritical mean-field is dissatisfied.
However, for Figs. \ref{MAP} (a), (b) and (c), it is difficult to distinguish which model is the best.
As we know, for an ideal model, the modified Arrott plot should display a series of parallel lines in high field region with the same slope which defined as $S(T)=dM^{1/\beta}/d(H/M)^{1/\gamma}$.
Thus, in an ideal model, the slopes of $M^{1/\beta}$ vs. $(H/M)^{1/\gamma}$ lines should exhibit the sample value as that at $T_{C}$.
We define the normalized slope ($NS$) as $NS= S(T)/S(T_{C})$.
For a most satisfied model, all $NS$ values should approach $'1'$ closely.
By this method, it enables us to distinguish the most suitable model by comparing the $NS$ with the ideal value of $'1'$ \cite{zhangPRB,FanPRB,zhangEPL}.
Plot of $NS$ vs. $T$ for the four different models is shown in Fig. \ref{ns}.
It can be seen that the $NS$ of 3D-Heisenberg model is close to $'1'$ mostly among these modes.

\begin{figure*}%[B]
\includegraphics[width=1.0\textwidth,angle=0]{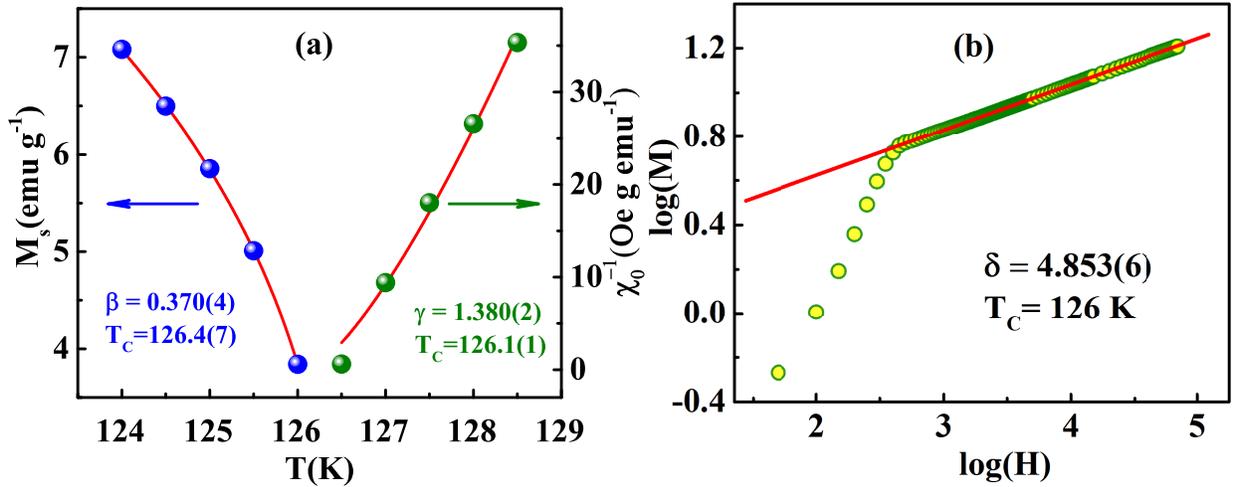}
\caption{(Color online) (a) The $M_{S}$ (left) and $\chi$$_{0}$$^{-1}$ (right) as a function of temperature for Cr$_{1/3}$NbS$_{2}$; (b) isothermal $M(H)$ at $T_{C}$ on log-log scale (red solid curves and line are fitted).} \label{exponents}
\end{figure*}

More accurate critical exponents $\beta$ and $\gamma$ can be obtained by the iteration method \cite{zhangleiPRB2015}, which are used to distinguish which universality class they belong to.
The initial parameters are based on those of Fig. \ref{MAP} (a) since its $NS$ values are close to $'1'$ mostly.
The linear extrapolation from high field region to the intercepts with the axes $M^{1/\beta}$ and $(H/M)^{1/\gamma}$ yields $M_{S}(T,0)$ and $\chi_{0}^{-1}(T,0)$ under zero field.
By fitting to Eqs. (\ref{1}) and (\ref{2}), a set of $\beta$ and $\gamma$ is obtained, which is used to reconstruct a new modified Arrott plot.
Subsequently, new $M_{S}(T,0)$ and $\chi_{0}^{-1}(T,0)$ are generated from the linear extrapolation from high field region.
By this way, another set of $\beta$ and $\gamma$ can be obtained.
This procedure is repeated until $\beta$ and $\gamma$ hardly change.
The final obtained critical exponents by the iteration method are independent on the initial parameters.
In this way, $\beta= 0.370(4)$ with $T_{C}=126.4(7)$ and $\gamma=1.380(2)$ with $T_{C}=126.1(1)$ are obtained for Cr$_{1/3}$NbS$_{2}$, as shown in Fig \ref{exponents} (a).
The obtained $T_{C}$ is consistent with that obtained by T. Miyadai $et$ $al$ \cite{Miyadai}.
Figure \ref{exponents} (b) shows the initial $M(H)$ at the critical temperature $T_{C}= 126$ K, with the inset plotted on a $\log-\log$ scale.
One can see that the $M(H)$ at $T_{C}$ exhibits a straight line on a $\log-\log$ scale when $H>H_{S}$.
Thus, $\delta=4.853(6)$ is determined in the high field region ($H>H_{S}$) based on Eq. (\ref{3}).
According to the statistical theory, these critical exponents should
fulfill the Widom scaling law \cite{Kadanoff}:
\begin{equation}
\delta=1+\frac{\gamma}{\beta}\label{4}
\end{equation}
According to the Widom scaling law $\delta=4.729(7)$ is calculated, which in agreement with that obtained from the experimental critical isothermal analysis.
The self-consistency of the critical exponents demonstrates that they are reliable and unambiguous.

The final critical exponents ($\beta= 0.370(4)$, $\gamma=1.380(2)$, $\delta=4.853(6)$) are in agreement with those recently reported by E. Clements $et$ $al.,$ \cite{Clements} which confirm the reliability of these obtained exponents.
Both sets of exponents from these two works are very close to the theoretical prediction of 3D-Heisenberg model ($\beta= 0.365$, $\gamma=1.386$, $\delta=4.80$), which suggest a short-rang magnetic coupling of Cr-Cr \cite{Kaul,Clements}.
Although the structure of Cr$_{1/3}$NbS$_{2}$ is two-dimensional, its magnetic coupling is of a three-dimensional type.
This indicates that the magnetic interactions of Cr-Cr are coupled not only within the $ab$-plane but also between the inter-layers along the $c$-axis.
As we know, the DM interaction plays an important role in the magnetic structure of Cr$_{1/3}$NbS$_{2}$, which makes the spins exhibit spiral ordering.
However, the DM interaction has rarely effect on the critical exponents due to two aspects.
The DM interaction is much weaker than the ferromagnetic exchange, which is only $1\sim2$ orders of magnitude of the ferromagnetic exchange \cite{Wilhelm}.
Thus, the critical exponents are mainly determined by the magnetic exchange interaction.
In fact, in the systems with DM interaction such as FeGe, MnSi, Cu$_{2}$OSeO$_{3}$, Fe$_{0.8}$Co$_{0.2}$Si, their critical exponents are either rarely influenced by the DM interaction \cite{Wilhelm,zhangleiPRB2015,Zivkovic,Jiang}.

\begin{figure}%[B]
\includegraphics[width=1.0\textwidth,angle=0]{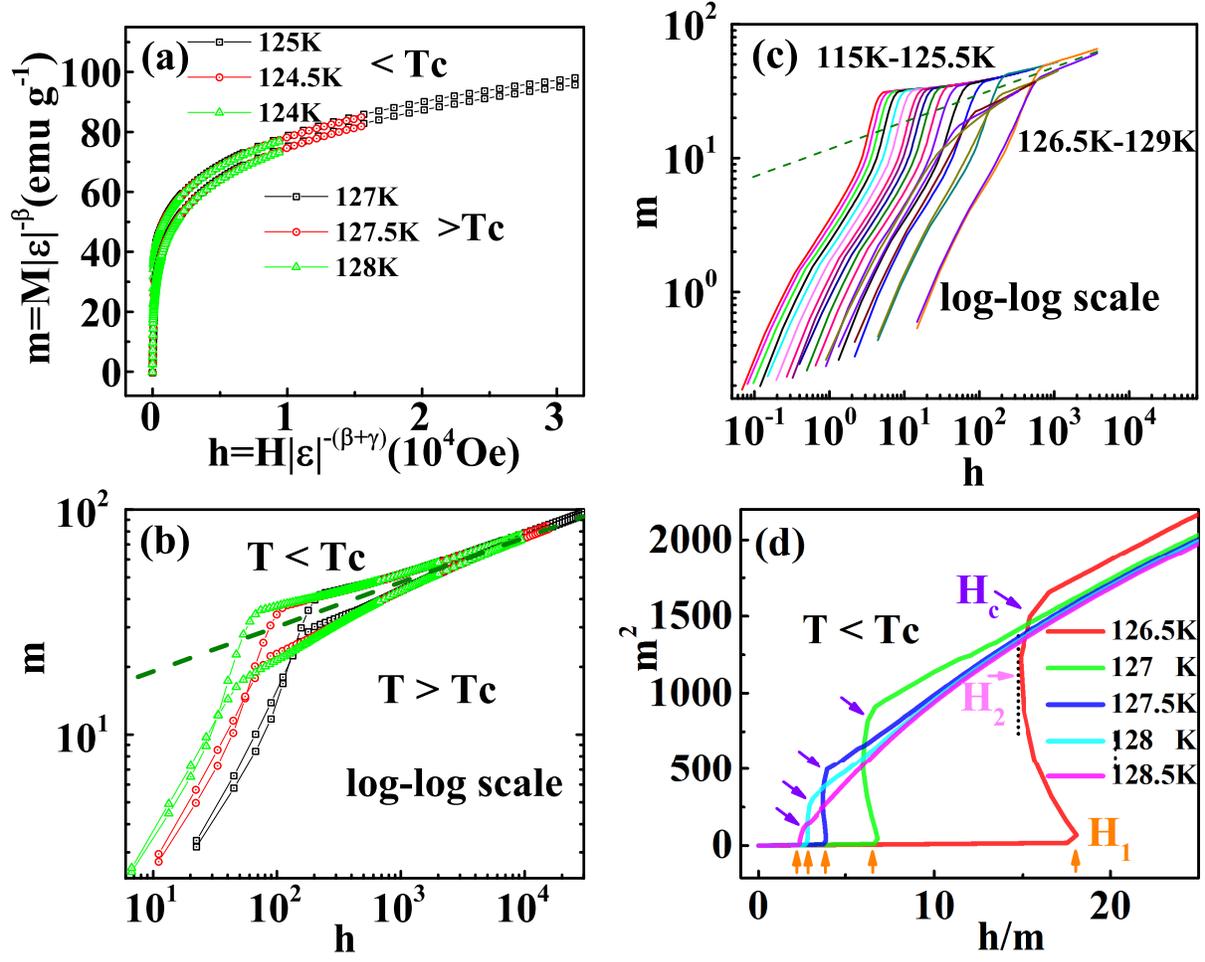}
\caption{(Color online) (a) Scaling plot of $m$ vs. $h$ around $T_{C}$ for $M-H$ at typical temperature; (b) typical $m$ vs. $h$ curves on log-log scale; (c) the magnification of $m$ vs. $h$ in low field region; (d) the re-scaling of $m^{2}$ vs. $h/m$ ($H_{1}$ is the critical field from HM to CSL-1; $H_{2}$ is the critical field from CSL-1 to CSL-2; and $H_{C}$ is the critical field from CSL-2 to FFM).}
\label{scaling}
\end{figure}

For a homogeneous magnet, M. E. Fisher $et$
$al.$ theoretically treated the magnetic ordering as
attractive interaction of spins, where a renormalization group
theory analysis suggests the exchange distance $\emph{J}(r)$ decays with distance $r$
as \cite{Fisher,Ghosh}:
\begin{equation}
\emph{J}(r)\approx r^{-(d+\sigma)}
\end{equation}
where $d$ is spatial-dimensionality, and $\sigma$ is a critical exponent which can calculated as \cite{Fisher,Fischer1}:
\begin{equation}
\gamma=1+\frac{4}{d}\frac{n+2}{n+8}\Delta\sigma+\frac{8(n+2)(n-4)}{d^{2}(n+8)^{2}}\left[1+\frac{2G(\frac{d}{2})(7n+20)}{(n-4)(n+8)}\right]
\Delta\sigma^{2} \label{Jr}
\end{equation}
where $\Delta\sigma=(\sigma-\frac{d}{2})$, $G(\frac{d}{2})=3-\frac{1}{4}(\frac{d}{2})^{2}$, $n$ is the spin-dimensionality.
For the 3D-Heisenberg model, there is $d=$ 3 and $n=$ 3.
It can be obtained that $\sigma=$ 1.932(6) for Cr$_{1/3}$NbS$_{2}$, which indicates that the magnetic interaction decays on spatial distance as $J(r)\sim r ^{-4.9}$ \cite{Fisher,zhangleiPRB2015}.
It has been suggested that if $J(r)$ decreases with distance $r$ faster than $r^{-5}$, the Heisenberg exponents are valid for a three-dimensional isotropic ferromagnet \cite{Fisher}.
As we know, these critical exponents should follow the scaling equations.
Defining the renormalized magnetization $m\equiv\varepsilon^{-\beta}M(H,\varepsilon$)
and the renormalized field $h\equiv H\varepsilon^{-(\beta+\gamma)}$, in the asymptotic critical region the scaling equations can be written as \cite{Stanley}:
\begin{equation}
m=f_{\pm}(h)\label{CF}
\end{equation}
where $f_{\pm}$ are regular functions denoted as $f_{+}$ for
$T>T_{C}$ and $f_{-}$ for $T<T_{C}$. The scaling equations indicate
that $m$ vs. $h$ should form two universal branches for $T>T_{C}$ and
$T<T_{C}$ respectively, even those in low field region \cite{Khan,Phan}.
However, if a field-induced phase transition occurs, the scaling becomes divergent at the boundary between the phases.
The divergence of the scaling curves supplies a method to distinguish the different phases \cite{zhangleiPRB2015}.
Based on the scaling equation, the isothermal magnetization around the
critical temperatures for Cr$_{1/3}$NbS$_{2}$ are replotted in Fig. \ref{scaling}
(a), and the typical $m$ vs. $h$ curves are shown on log-log scale in the Fig. \ref{scaling} (b) to magnify those in low field region.
It can be seen that experimental data collapse into two universal curves, except the $M-T-H$ in the low field region.
Especially, the $m$ vs. $h$ on log-log scale is depicted in Fig. \ref{scaling} (c), where the magnetic transitions in low field region are displayed.
Furthermore, $m^{2}$ vs. $h/m$ curves clearly distinguish the different phases in low field region, as shown in Fig. \ref{scaling} (d).
Three regions separated by two transitions are clearly seen on $m^{2}$ vs. $h/m$ curves, which are corresponding to the HM, CSL, and FFM states.
In very low field region, a plateform is found on $m^{2}$ vs. $h/m$ curves, which is corresponding to the HM state.
With the increase of field, the abrupt increase region of $m^{2}$ vs. $h/m$ is attributed to the CSL transition, where the critical field from HM to CSL is defined as $H_{1}$.
When $H$ exceeds $H_{S}$ ($H_{S}$ is the critical field from CSL to FFM), the CSL is completely polarized into FFM state.
In the CSL region, the state can be further identified by the slope ($S$) of $m^{2}$ vs. $h/m$.
In fact, the transition from CSL to FFM phase is a procedure of competition between CSL and FFM.
In lower field region, the CSL is dominant where the magnetic solitions are separated by the FM region (CSL-1 state) \cite{Tsuruta}.
In higher field region, the FFM phase becomes dominant in the system, where the whole magnetization is very close to the FM behavior (CSL-2 state) \cite{Tsuruta}.
It can be seen that the change from negative to positive occurs to $S$ of $m^{2}$ vs. $h/m$ in the CSL region, where the critical field from CSL-1 to CSL-2 is defined as $H_{2}$.
As we know, the FFM behavior exhibits positive $S$ ($S>0$) \cite{Banerjee}.
Thus, the positive $S$ in CSL region corresponds to CSL-2 (with large FM array) which approaches the FFM behavior with $S>0$.
Alternatively, the region of $S<0$ is corresponding to CSL-1 (dominant helical texture with poor FM array) \cite{Tsuruta}.
By this method, we can distinguish the CSL-1 from CSL-2 states.

\begin{figure}%[B]
\includegraphics[width=1.0\textwidth,angle=0]{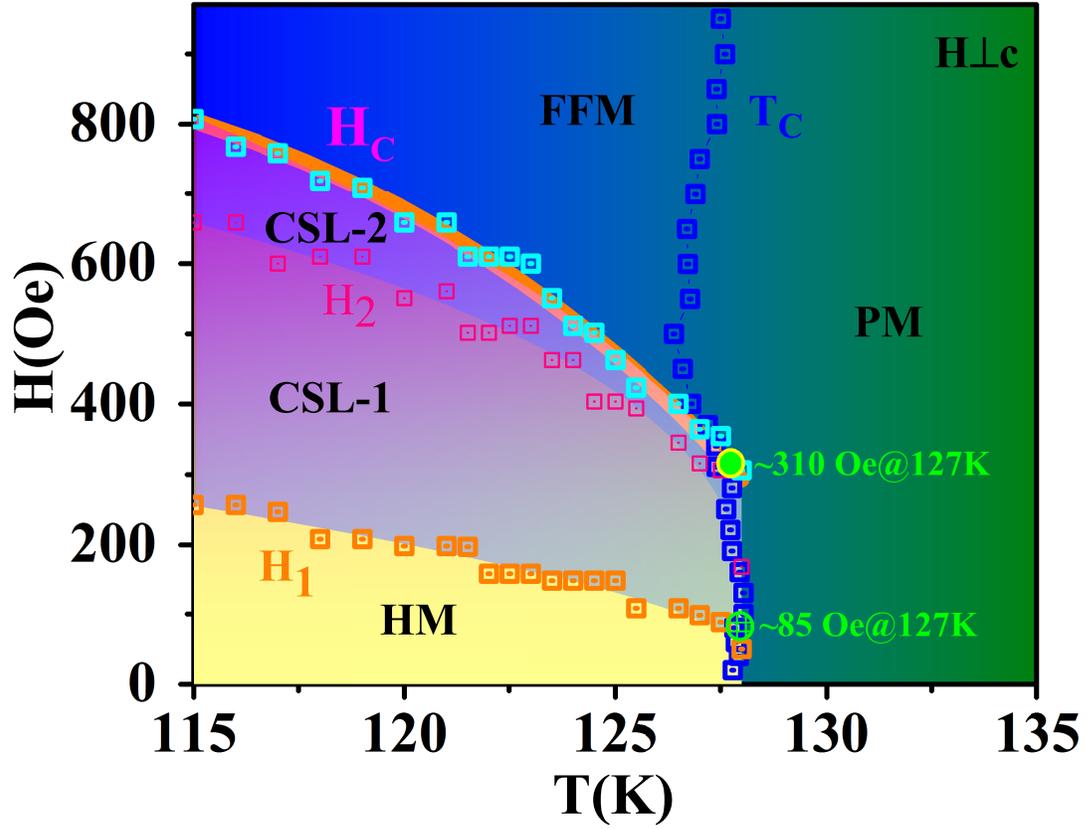}
\caption{(Color online) The $H-T$ phase diagram in the vicinity of the phase transition obtained by scaling of the $M-T-H$ for Cr$_{1/3}$NbS$_{2}$ with $H\perp c$ (HM corresponds to the helimagnetism; CSL is the chiral magnetic soliton phase; PM is the paramagnetic state; and FMM represents the forced ferromagnetic state).}
\label{phase}
\end{figure}

Based on the scaling of the $M-T-H$, the $H-T$ phase diagram with $H\perp c$ is constructed in Fig. \ref{phase}.
Since the critical scaling is restrict within the critical temperature region ($|(T-T_{C})/T_{C}|\leq 0.1$), only the phase diagram around $T_{C}$ is mapped.
It can be seen that Cr$_{1/3}$NbS$_{2}$ exhibits a HM ground state below $T_{C}\sim$ 127 K.
When field is applied, the HM state is modulated to CSL-1, and subsequently changes to CSL-2 state.
Finally, the CSL-2 state is polarized into FFM phase when $H$ exceeds $H_{C}$.
Two critical points can be determined on the $H-T$ phase diagram, as shown in Fig. \ref{phase}.
In fact, in the monoaxial chiral helimagnetic system, two tricritical points have been predicted theoretically by V. Laliena $et$ $al.$ when the field is applied along variable directions \cite{Laliena2016}.
However, in another work, they suggests a tricritical point and a zero-field critical one in monoanixal helimagnet when field is applied along the $c$-axis \cite{Laliena}.
The zero-field critical point locates at boundary from HM to PM phases \cite{Laliena}.
Therefore, in present case, one critical point corresponds to a tricritical one, which locates at the intersection between the CSL, FFM, and PM phases ($\sim 310$ Oe at 127 K).
The other one is determined as zero-field critical point, which is situated at the boundaries between HM, CSL and PM phases ($\sim 85$ Oe at 127 K).

\section{Conclusion}

In summary, the magnetism of the single crystal Cr$_{1/3}$NbS$_{2}$ is investigated.
The magnetization displays strong magnetic anisotropy when the external field is applied parallel and perpendicularly to the $c$-axis ($H<H_{S}$).
The $M(H)$ curve with $H\perp c$ displays a loop on the boundary of the phase transition from incommensurate CSL to commensurate FFM, which suggests that the CSL phase is a magnetic metastable.
The critical exponents of Cr$_{1/3}$NbS$_{2}$ are obtained as $\beta=$ 0.370(4), $\gamma=$ 1.380(2), and $\delta=$ 4.853(6), which are close to the theoretical prediction of 3D-Heisenberg model.
The critical behavior indicates the magnetic coupling is of a short-range type.
Based on the scaling equation, the $H-T$ phase diagram around $T_{C}$ is constructed, where two critical points are determined.
One critical point is a tricritical one, which locates at the intersection between the CSL, FFM, and PM states.
The other critical point is a zero-field critical one, which is situated at the intersection between CSL, HM and PM states.

\section{Acknowledgements}

This work is supported by the National Key R\&D Program of China (Grant No. 2017YFA0303201) and the National Natural Science Foundation of China (Grant Nos. 11574322, U1732276, 11474290, and 11574288).
D. S. and D. G. M. acknowledge support from the National Science Foundation under grant DMR-1410428.

%The authors thank Dr. Jiyu Fan and Charles Hwang for their fruitful discussion and review of this paper.

%Grant Nos. 11574322, U1332140 (L. Zhang)
%U1532267, (C. J. Zhang)
%11574288 (L. Pi)
%11174291 and U1532153 (Z. Qu)
%11474290 (xd zhu)
%11204288 M.Ge
%11574288 L. Pi
%11474290 HF.du
%11104281 HF. du

%\section*{Author Contributions}
%
%H. H. and L. Z. conducted all of the experiments and wrote the paper.
%H. F. D, M. G. and L. S. L. performed the magnetic measurements.
%C. J. Z., L. P. and Y. H. Z. analyzed the experimental results.
%
%\section*{Additional Information}
%
%\textbf{Competing financial interests:} The authors declare no competing financial interests.

\end{document}